\documentclass[aps,prb,twocolumn,showpacs]{revtex4-1}
\usepackage{graphicx}
\usepackage{dcolumn}
\usepackage{bm}
\usepackage{amsmath}
\usepackage{tabularx}
\usepackage{float}
\usepackage{pifont}
\usepackage{rotating}  
\usepackage{hyperref}  
\usepackage{multirow}
\newfloat{widefig}{thp}{lop}
\begin{document}
\title{Site occupancies and their effects on the physical properties of 
spinel $Co\left(Cr_{1-x}Fe_{x} \right)_{2}O_{4}$: an {\it ab initio} study}
\author{Debashish Das} %
\affiliation{Department of Physics, Indian Institute of Technology
Guwahati, Guwahati, Assam 781039, India} %

\author{ Subhradip Ghosh}
\affiliation{Department of Physics, Indian Institute of Technology
Guwahati, Guwahati, Assam 781039, India} %

\date{\today}

\begin{abstract}
 Recent experimental studies on Fe substituted spinel CoCr$_{2}$O$_{4}$ have discovered multiple functional properties in the system such as temperature and composition dependent magnetic compensation, tunable exchange bias and magnetostriction. These properties are attributed to the renormalisation of the inter-atomic magnetic exchange interactions arising due to the non-regular site occupancies of the magnetic cations in the system. The site occupancy patterns and the resulting modifications in various competing magnetic exchange interactions can explain the evolution of the collinear magnetic structure when Fe is gradually substituted in CoCr$_{2}$O$_{4}$ having a non-collinear magnetic structure. In this work, we perform {\it ab initio} electronic structure calculations by DFT+U method and combine with a generalised thermodynamic model [J. Phys. Condens. Matter 29, 055895 (2017)] to compute the site occupancy patterns of the magnetic cations, the structural properties and the magnetic exchange interactions of Co$\left(Cr_{1-x}Fe_{x} \right)_{2}$O$_{4}$ for the entire composition range $0<x<1$. We find that the substituting Fe atoms prefer to occupy the tetrahedral sites of the spinel structure for the entire range of $x$, in agreement with the experimental inferences. Our results on the variations of the structural parameters with compositions agree very well with the experiments. By computing the variations of the various inter-atomic magnetic exchange interactions, we provide a microscopic picture of the evolution of a collinear structure from a non-collinear one due to substitution of Fe in CoCr$_{2}$O$_{4}$. The computed results are analysed in terms of the elements of the crystal field theory, and the features in the atoms and orbital-projected densities of states. The results and analysis presented in this work is the first comprehensive study on this system which would help understanding the complexities associated with the site occupancies, the electronic structures and the magnetic interactions in this multi-functional material.     
\end{abstract}

\maketitle
\section{Introduction}
 The spinel oxide CoCr$_{2}$O$_{4}$ has been the subject of intense investigations  since the discovery of multiferrocity in this material \cite{polarization}. The system has a rich magnetic phase diagram due to the strong coupling between the magnetic cations at different sites \cite{cco1,cco2}. The conical spin structure in this material, at low temperature, responsible for it's multiferroic nature, is understood to be an artefact of comparable B-B and A-B magnetic exchange interactions \cite{menyuk,lkdm}; A sites are the tetrahedral sites and B sites are the octahedral sites in an AB$_{2}$O$_{4}$ spinel compound. The origin of the non-collinear spin structure in CoCr$_{2}$O$_{4}$ is further examined by {\it ab initio} density functional theory based calculations \cite{bs,ederer}. The results confirm the picture of the competing Co-Cr and Cr-Cr exchange interactions driving the system towards the non-collinear spin structure and that the experimentally observed conical structure is due to the nesting features in the Fermi surface of this system.

Since the origin of the non-collinear spin structure is in the delicate balance of magnetic exchange interactions between various magnetic cations (Co and Cr in this case) occupying different crystallographic sites in CoCr$_{2}$O$_{4}$, it is only natural to investigate the influences of the variations in the occupancies at the A and B sub-lattices on the ferroelectric polarisation and on the physical properties, in general. To this effect, few  studies by substituting Cr with Fe \cite{fe1,fe2,fe3,fe4,fe5} and with Mn \cite{mn} have been done recently. In all such studies, an intricate relationship between the structural properties, the magnetic properties and the occupancies at different cationic sites were observed. Structural and magnetic measurements for Mn substituted CoCr$_{2}$O$_{4}$ (Co$\left(Cr_{1-x}Mn_{x} \right)_{2}$O$_{4}$) led the experimentalists to conclude that depending on the concentration $x$ of the substituting Mn, the occupancies at the A and B sites change considerably, leading to phenomena such as magnetic compensation \cite{mn}. The experimental observations and the inferences were complemented by the density functional theory based calculations, in conjunction with a thermodynamic model \cite{dd1}. In this first-principles calculation, the site occupancies at A and B were obtained from a thermodynamic analysis, and the observed anomalous behaviour of the total magnetisation was analysed based upon that. The substitution of Fe at the Cr site is expected to provide more interesting aspects of the inter-relations between the site occupancies and the magnetic interactions. The end compounds CoCr$_{2}$O$_{4}$ and CoFe$_{2}$O$_{4}$ have very different site occupancy patterns as well as magnetic structures, which was not the case for the end compounds in case of Mn-substituted CoCr$_{2}$O$_{4}$. CoCr$_{2}$O$_{4}$ is considered a "normal" spinel where the A sites are occupied by Co$^{2+}$ ions and B sites by Cr$^{3+}$ ions. On the contrary, CoFe$_{2}$O$_{4}$ is considered an "inverse" spinel where the A sites are primarily occupied by the Fe atoms while the Co atoms occupy one of the B sites. The "degree of inversion" or the "degree of cation disorder" is decided by a single parameter $y$ which denotes the concentration of B site atoms occupying the A sites. Thus, when $y=0$, the system is considered "normal". The complete "inverse" structure is the one with $y=1$ when one of the Fe atoms completely occupy the A sites. For CoFe$_{2}$O$_{4}$, the degree of cation disorder is found to be the decisive factor for it's ground state. While one work \cite{cfo1} reported the ground state of this system to be insulating, another group \cite{cfo2} reported it to be a half-metal. The calculations by Ganguly {\it et al}\cite{bs} conclusively showed that the cation disorder at A and B sites, irrespective of the degree decided by $y$, drives the system towards half-metallicity due to the significant changes in the electronic structure. Also, the magnetic transition temperature of CoFe$_{2}$O$_{4}$ is quite high (860 K) in comparison to CoCr$_{2}$O$_{4}$. Recent {\it ab initio} calculations of exchange interactions in complete "inverse" CoFe$_{2}$O$_{4}$\cite{dd2} showed that the large value of the magnetic transition temperature is due to a collinear spin structure which is driven by overwhelmingly dominating A-B interactions. Thus, when Cr is gradually substituted by Fe, to make a systematic transition towards CoFe$_{2}$O$_{4}$ from CoCr$_{2}$O$_{4}$, two different factors, the degree of cation disorder and the magnetic exchange interactions, would play pivotal roles.

Recent experimental attempts of Fe substitution in CoCr$_{2}$O$_{4}$ has been remarkable. Magnetic measurements on 10$\%$ and 15$\%$ Fe substituted CoCr$_{2}$O$_{4}$ reported a magnetisation reversal and a sizeable exchange bias around a critical temperature, accompanied by non-monotonic changes in the local moments \cite{fe2,fe3,fe4,fe5}. They attributed these phenomena to the magneto-structural changes arising due to the cation disorder and subsequent spin re-orientations at the cationic sub-lattices, due to the presence of Fe substituting Cr. Magnetic and specific heat measurements \cite{fe1} for Co$\left(Cr_{1-x}Fe_{x} \right)_{2}$O$_{4}$, $0\leq x \leq 0.5$ scanned a larger composition regime and concluded that the magnetisation reversal can be understood in terms of role reversals of the magnetic contributors at different sub-lattices, the origin of which lies in the changes in the sub-lattice occupancies of the Fe atoms. Their specific heat versus temperature results indicated tat the occupancies of the Fe atoms at different sub-lattices was the reason behind the suppression of the conical spin structure and the emergence of a collinear magnetic structure as Fe concentration in the system increases. Density functional theory based calculations for $6.25 \%$ and $12.5 \%$ of Fe substitution in CoCr$_{2}$O$_{4}$ provided important insights into the impacts of cation disorder on the local structural properties, the electronic structure and the magnetic exchange splitting \cite{dd3}. It was found that the electron-electron strong correlation plays important role in deciding the spin states of the magnetic atoms in this system, and that the substitution of Fe at the tetrahedral A sites, instead of the octahedral B sites introduces significant local distortions which modify the electronic structure at the cationic sites. Total energy calculations, even at such low values of $x$, clearly showed the tendency of the system towards "inversion", in confirmation with the picture provided by the experimentalists. The calculations of the magnetic exchange parameters for $0\leq x \leq 0.1$ \cite{bs} for the "completely inverted" and "normal" structures showed that the A-B exchange interactions gain significance over the B-B ones as $x$ increases, thus predicting that the system will be driven towards collinearity by the exchange interactions between Fe atoms at A sites and Co atoms at B sites.

Two things were hitherto unavailable in these investigations: (i) the computations of the "degree of cation disorder" $y$ at different values of $x$ in Co$\left(Cr_{1-x}Fe_{x} \right)_{2}$O$_{4}$ so that the ideas about the site-occupancies at the operational temperatures of the experiments can be obtained and (ii) a systematic investigation into the effects of site preferences of the cations on the physical properties of this system for the complete composition range ($0\leq x \leq 1$) within same approximations and a single methodological framework so that a consistent trend and thus, the understanding of the physics associated with the system, can be achieved. In this work, we have addressed these two shortcomings of the available works on this system by computing the "inversion parameter" $y$, and investigating it's effects on the structural properties and the magnetic exchange parameters as $x$ varies. We have analysed our results by computing the dependencies of the electronic structures on the site occupancies and provided a microscopic picture of the competing exchange interactions that lead the system from a non-collinear magnetic structure on one end to a collinear structure on the other. In next section, we briefly discuss the models and other computational details used in this work. The results are presented and analysed in the next section followed by the conclusions.

\section{Details of calculations}
We have modelled Co$\left(Cr_{1-x}Fe_{x} \right)_{2}$O$_{4}$ system for various values of $x$ and various degrees of cation disorder $y$ according to the scheme presented in Reference \cite{suppl}. The details of calculations of the thermodynamic free energy, and the procedure of estimating the equilibrium $y_{0}$ for a given temperature $T$ and Fe concentration $x$  have also been discussed in Reference \cite{suppl}. 
As done in Reference \cite{suppl},the magnetic configuration in all cases have been taken to be Neel 
configuration, with
spins of $A$ and $B$ sub-lattices anti-aligned. Before fixing Neel configuration as the magnetic configuration, we have done
several calculations with different spin configurations at different sub-lattices. In almost all
cases, the Neel configuration came out to be energetically lowest. In cases where it was
not, the lowest energy spin configurations were lower by less than 0.1 meV per atom. This,
thus, further justifies consideration of Neel configuration for all $x$ and $y$. 
The total energies, the structural parameters, the magnetic moments and the electronic
structures were calculated by the DFT+U \cite{dft-u} method using Projector Augmented
Wave (PAW) \cite{paw} basis set as implemented in VASP \cite{vasp} code. The effects
of electron localisation were addressed by the approach of Dudarev {\it et al} \cite{dudarev}.
The Hund's coupling parameter $J$ was taken to be $1$ eV, while the Coulomb parameter
$U$ was taken to be $5$ eV for Co and $3$ eV for Cr and Fe. A plane wave cut-off
of $550$ eV and a $5 \times 5 \times 5$ mesh centred at $\Gamma$ point for Brillouin zone
integrations have been used throughout with the only exception for $x=0.0625$ where a 
$2 \times 2 \times 2$ mesh was enough to achieve an energy convergence of $10^{-7}$ eV.
Force convergences of $10^{-4}$ eV/$\AA$ were ensured during structural relaxations. 
 
\section{Results and Discussions}
\subsection{Temperature and concentration dependences of degree of cation disorder
$y$}
\begin{figure}[ht]
\includegraphics[width=8cm,height=8cm]{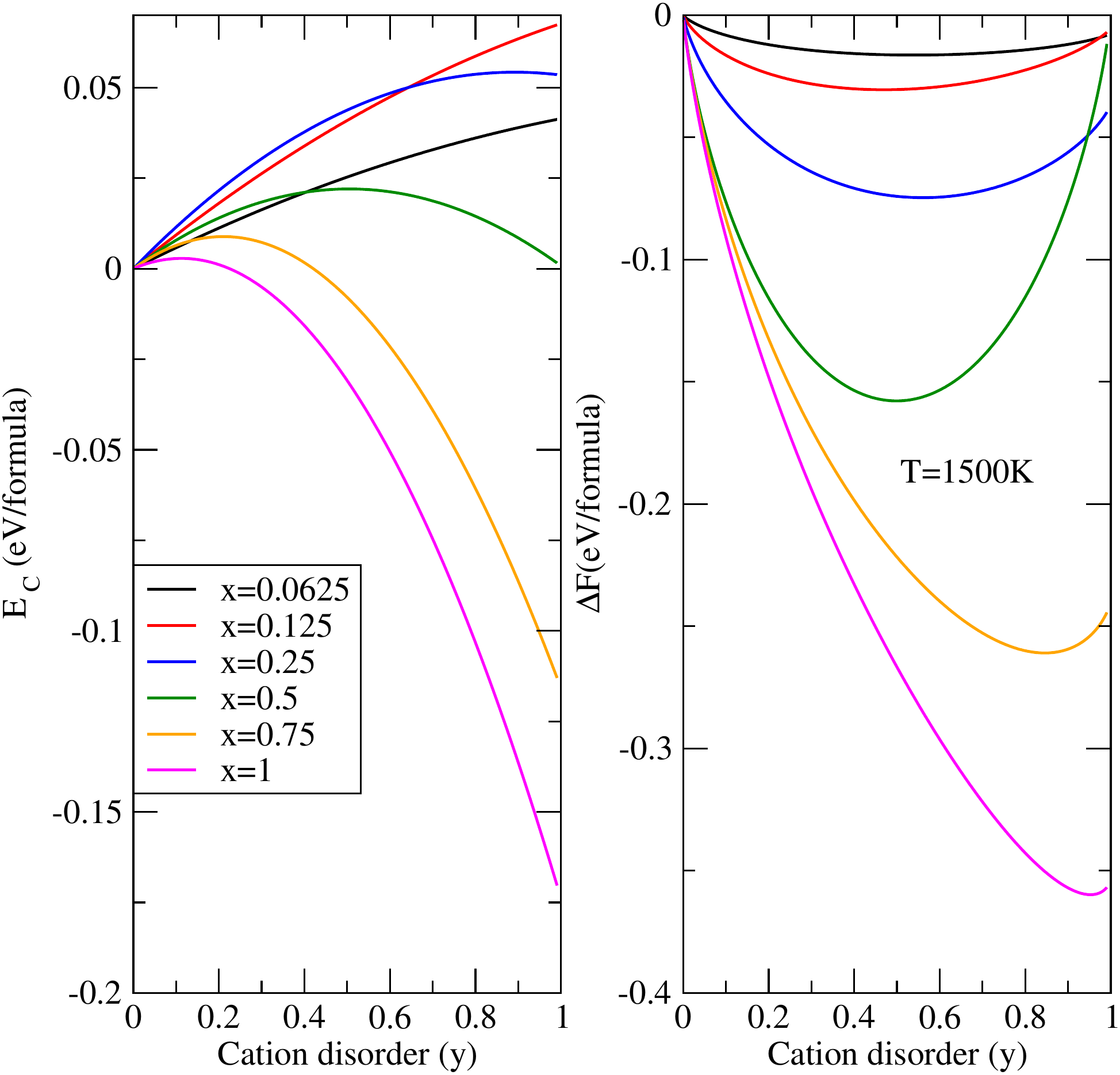}
\caption{\label{fig1} Variations of the cation disorder energy($E_c$)
(left panel) and the configurational free energy ($\Delta F$)(right panel)
 with degree of cation disorder $y$ of $Co\left(Cr_{1-x}Fe_{x} \right)_{2}O_{4}$, 
for different values of $x$, the Fe
concentration, at 1500 K, the annealing temperature of the experiment \cite{fe1,fe4}.
The equilibrium inversion parameter ($y_0$) at a given $T$ and
for a given $x$ is obtained from the minima of $\Delta F$.}
\end{figure}

\begin{figure}[ht]
\includegraphics[width=8cm,height=8cm]{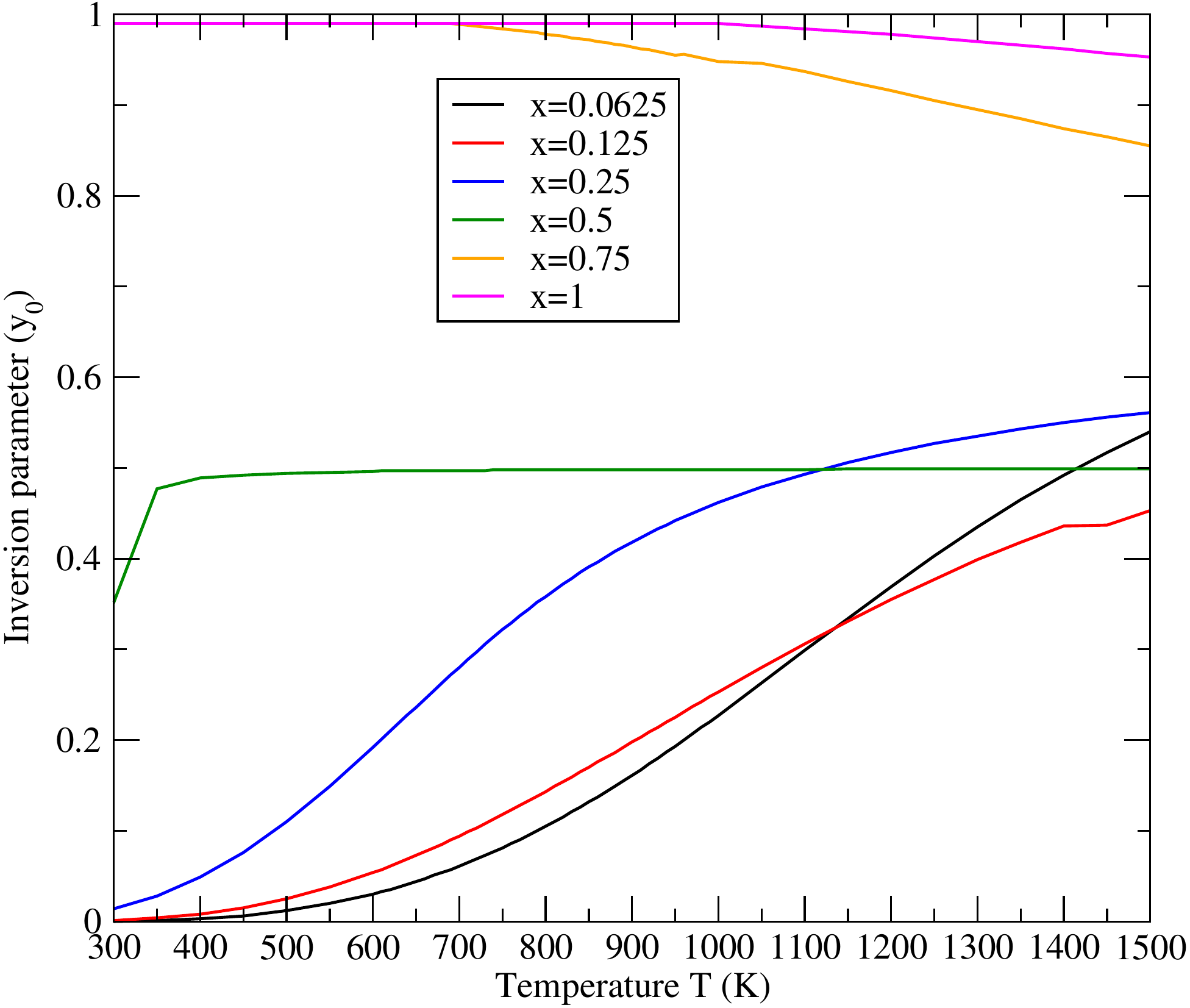}
\caption{\label{fig2} Temperature dependence  of the  
equilibrium inversion parameter ($y_0$) in $Co\left(Cr_{1-x}Fe_{x} \right)_{2}O_{4}$
for different $x$ for the temperature ranging from room temperature to the
annealing temperature of the experiment \cite{fe1,fe4}.}
\end{figure}

In Fig. \ref{fig1} we show the dependences of the cation disorder energy $E_{c}$ and the
configurational free energy of cation disorder $\Delta F$ on $y$ for different $Fe$
concentration $x$, at a temperature $1500$ K which
is close to the annealing temperature of $1473-1573$ K reported in the experiment
\cite{fe1,fe4}. Unlike Mn-substituted CoCr$_{2}$O$_{4}$, in this case, we find that the cation disorder is zero for smaller values of $x$ only, if we do not
consider the configurational entropy. When Fe substitution is $50 \%$, the configurational energy is nearly identical for the "normal" and the complete "inverse" structures. For higher $x$ values, the lowest energy state is the complete "inverse" one. T. 
Inclusion of the entropy term stabilises states with large values of $y$ for all compositions. The results at $1500$ K suggest that upto $x=0.5$, $y_{0}$, the equilibrium value of $y$ is nearly 0.5 for all $x$ values, which means that nearly $50 \%$ of the substituted Fe prefer to occupy the tetrahedral sites even when it's concentration in CoCr$_{2}$O$_{4}$ is small, thus making these systems close to a "half-inverse" configuration. For $x=0.75$ and $1$, the results suggest that at $1500$K, the high Fe concentration compounds prefer to be nearly "complete inverse" as $y_{0}$ is close to 0.9 for these two compositions. The quantitative variations of 
$y_{0}$ with temperature $T$, shown in Fig. \ref{fig2} shows 
this qualitative behaviour clearly.  The results suggest that the cation disorder becomes more and more robust as the Fe concentration increases. For example, at $x=0.5$, the degree of cation disorder, denoted by $y_{0}$ stays at $0.5$ almost up to room temperature, while for lower values of $x$, $y_{0}$ varies substantially with temperature. For higher values of $x$, that is, when Fe concentration in the system is more than $50 \%$, the system is nearly "complete inverse" for the entire temperature range. This result is in good agreement with the site occupancy patterns in this system, suggested by the experimentalists. The XRD and EXAFS experiments by Zhang. {\it et al} \cite{fe1} suggested that except for $x=0-0.05$, the substituted Fe atoms dominantly distribute themselves on the A sites. They explained the phenomenon of magnetic compensation for $x=0.05$ and the absence of it for higher $x$ values based upon this site occupancy pattern of Fe.  

\subsection{Structural parameters and their variations with $x$ and $y$}
 \begin{table}[bp]
    \caption{\label{table1} Calculated cation-anion bond distances (in $\AA$) in $Co\left(Cr_{1-x}Fe_{x}
    \right)_{2}O_{4}$ for different $x$ and $y$ values.  }
    \vspace {2mm}
    \centering
        \begin{tabular}{c|c|c|c|c|c|c} \hline \hline
         &       & \multicolumn{2}{|c|}{Tetrahedral site} & \multicolumn{3}{|c}{Octahedral site}       \\ \hline
x        & y     & Co-O    &  Fe-O    &    Cr-O  &  Fe-O       & Co-O       \\ \hline
0        & 0      & 1.99   &  -        &   2.02   &  -          & -            \\ \hline
         & 0      & 1.99   &  -        &   2.02   & 2.04        & -               \\
0.0625   & 0.5    & 1.99   &  1.94     &   2.02   & 2.03        & 2.08         \\
         & 1      & 1.99   &  1.92     &   2.02   &   -         & 2.08       \\ \hline  

         & 0      &  1.98  &  -        &   2.02   & 2.03        & -            \\
0.125    & 0.5    &  1.98  &  1.92     &   2.02   & 2.02        & 2.08         \\
         & 1      &  1.99  &  1.93     &   2.01   &  -          & 2.09             \\ \hline           

         & 0      &  1.98  &  -        &   2.02   &  2.03       & -            \\
0.25     & 0.5    &  1.99  &  1.92     &   2.02   &  2.03       & 2.08         \\
         & 1      &  1.98  &  1.93     &   2.01    &  -          & 2.08        \\ \hline 
         
         & 0      &  1.98  &  -        &   2.02   & 2.04        & -            \\
0.5      & 0.5    &  1.97  &  1.93     &   2.02   & 2.03        & 2.09            \\
         & 1      &  -     &  1.92     &   2.01   & -           &  2.08         \\ \hline           

         & 0      & 1.98   &  -        &   2.03   & 2.04        & -            \\
0.75     & 0.5    & 1.97   &  1.92     &   2.02   & 2.04        & 2.08        \\
         & 1      & -      &  1.90     &   2.01   & 2.03        & 2.08    \\ \hline           
                  
         & 0      & 1.98   &  -        &  -       & 2.05        & -            \\
1.0      & 0.5    & 1.97   &  1.91     &  -       & 2.04        & 2.10           \\
         & 1      &  -     &  1.91     &  -       & 2.04        & 2.08   \\ \hline  \hline         
                       
         \end{tabular}      
\end{table}

\begin{table}[ht]
    \caption{\label{table2} Calculated structural parameters of $Co\left(Cr_{1-x}Fe_{x}
    \right)_{2}O_{4}$ for different $x$ and $y$ values. The lattice constant $a$ is in
    $\AA$. $u$ is the oxygen parameter. }

\begin{center}
\begin{tabular}
          { l@{\hspace{0.5cm}}  l@{\hspace{0.5cm}} l@{\hspace{0.5cm}} r@{\hspace{0.5cm}}c@{\hspace{0.5cm}} } \hline \hline
$x$        & $y$     & $a$   &    $u$        \\ \hline
0          & 0      & 8.430  &  0.262          \\ \hline
           & 0      & 8.434  &  0.262         \\
0.0625     & 0.5    & 8.432  &  0.262       \\
           & 1      & 8.430  &  0.261        \\ \hline  
           & 0      & 8.438  &  0.261         \\
0.125      & 0.5    & 8.433  &  0.260       \\
           & 1      & 8.428  &  0.260        \\ \hline          
           & 0      & 8.454  &  0.260         \\
0.25       & 0.5    & 8.437  &  0.260       \\
           & 1      & 8.427  &  0.260        \\ \hline 
           & 0      & 8.467  &  0.260         \\
0.5        & 0.5    & 8.452  &  0.258       \\
           & 1      & 8.416  &  0.257        \\ \hline  
           & 0      & 8.482  &  0.260         \\
0.75       & 0.5    & 8.453  &  0.258       \\
           & 1      & 8.433  &  0.256        \\ \hline 
           & 0      & 8.498  &  0.259         \\
1          & 0.5    & 8.468  &  0.257       \\
           & 1      & 8.437  &  0.255        \\ \hline            
 \end{tabular}
       \end{center}
\end{table} 

\begin{figure}[!t]
\includegraphics[width=8.0cm,height=10.0cm]{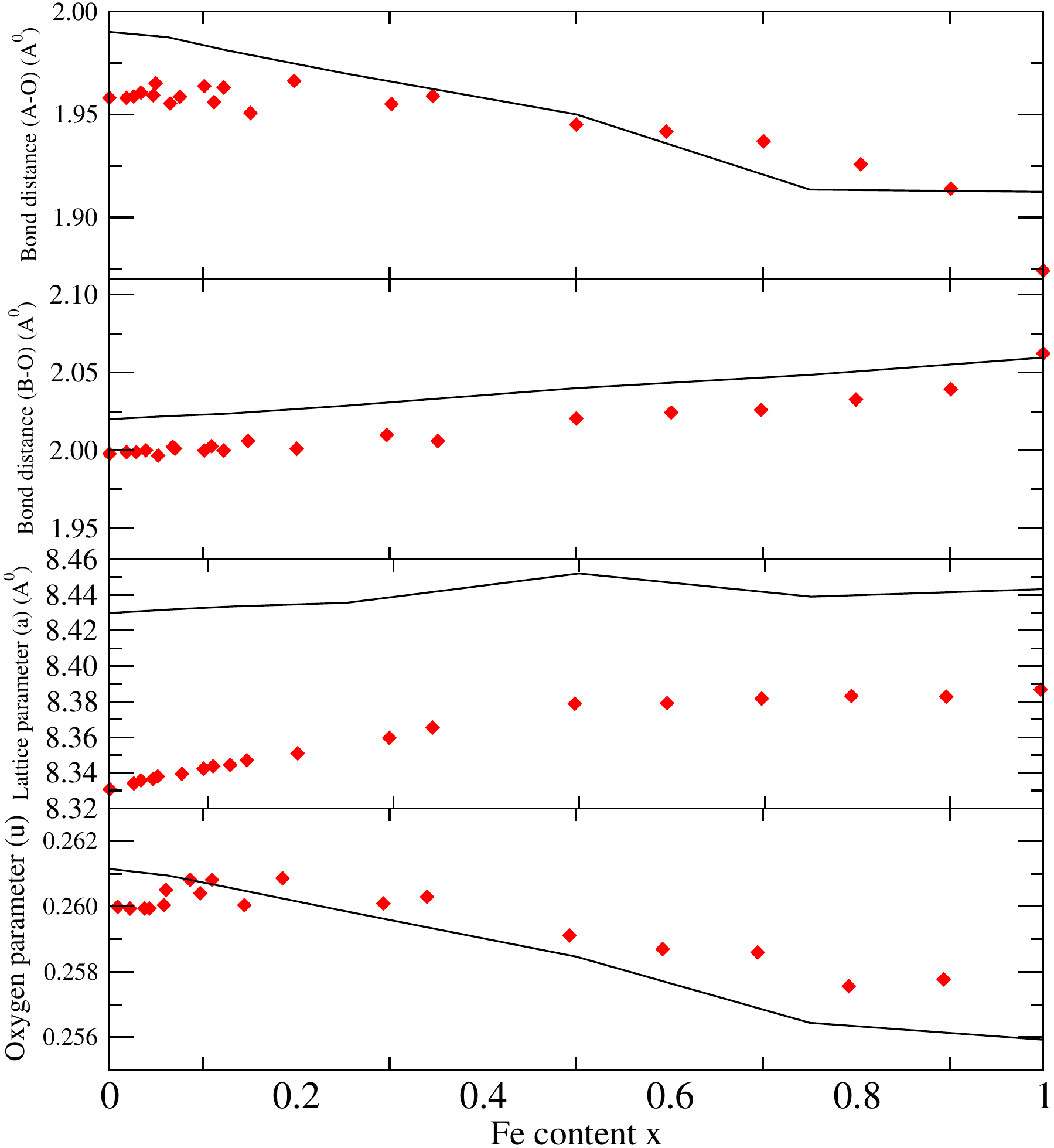}
\caption{\label{fig3} The calculated cation-anion bond distances ($A-O$ and $B-O$ in $\AA$), and the lattice parameters ($a$ in $\AA$ and the oxygen parameter $u$) as a function of Fe concentration $x$ for Co$\left(Cr_{1-x}Fe_{x} \right)_{2}$O$_{4}$. The Red circles are the experimental results \cite{fe4} plotted for comparison.}
\end{figure}
The anomalous change of slope in the variation of the lattice constant($a$) with $x$ obtained from the XRD pattern \cite{fe1} of this system first led to the belief that there can be a significant cation disorder in this system. The variations in the lattice constants for $x$ upto 0.05, obtained from their XRD measurements, showed a very slow increase while for the range $x=0.05-0.5$, the increase in the lattice constant with $x$ was rather rapid. The overall increase in the lattice constant was understandable for this system because the ionic radius of Fe$^{3+}$ in the octahedral (B-site) crystal field is 0.785 $\AA$, larger than that of Cr which is 0.755 $\AA$. However, Co$^{2+}$ has a much larger ionic radius of 0.885 $\AA$ in the octahedral crystal field than in the tetrahedral (A-site) crystal field where it's ionic radius is 0.72 $\AA$. In the tetrahedral crystal field, the ionic radius of Fe$^{3+}$ is only 0.63 $\AA$, much smaller than that of Co$^{2+}$.  Thus, if Fe ions mostly occupy the A sites and the Co atoms the B sites, the increase in the overall lattice constant would be rapid than if Fe had occupied the B sites, replacing the Cr atoms. The lattice constants obtained from the other XRD study \cite{fe4} by and large agreed with this explanation. However, they found that the rate of increase in the lattice constant slowed down for $x \geq 0.5$. They attributed this to the preferential occupation of octahedral B sites by the Fe atoms for higher Fe concentrations. Given these information from the experiments, we first discuss the variations in the structural parameters
with concentration of $Fe$ as well as with variations in the degree of cation disorder in order to understand the effects of Fe substitution on the structural parameters.
In Table \ref{table1}, we present various cation-anion bond distances at
sites of different symmetries  and their variations with $x$ and $y$. In Table \ref{table2},
we present the lattice constants $a$ along with the oxygen parameter $u$
for various $x$ and $y$. The lattice constants and the
cation-anion bond distances are obtained from the DFT+U
calculations. In Figure \ref{fig3}, we make a comparison between the experimental results \cite{fe4} and the results obtained from our calculations for the average cation-anion bond distances at different sites, and the lattice parameters.
The average cation-anion bond distances, for a given $x$, presented in figure \ref{fig3}, are obtained the following way: at a given value of $y$, the average cation-anion bond distance for a given site is obtained from the concentration averages of individual cation-anion bond distances. The value of $y_{0}$ for a particular $x$ is already known from Figure \ref{fig2}. The average cation-anion bond distance for the given $x$ is then obtained by interpolations of the average cation-anion bond distances for different $y$ values. The oxygen parameter $u$, is then obtained using these average values. The lattice constant for a given $x$ is also calculated in a similar way by interpolating the results for various $y$ values presented in Table \ref{table2}.  From Figure \ref{fig3}, we see that the qualitative agreement between the experimental results and our calculated results are quite good for the cation-anion bond distances and the oxygen parameter $u$. The disagreements are mostly in the range of $0<x<0.0625$ where we do not have any results from our calculations. The significant disagreement between the experimental results and the calculated values, however, is observed for the lattice constant $a$. While the experimental results suggest an increase of about $0.05\%$ as $x$ varies from $0$ to $1$, our calculations are unable to capture this small change. In order to capture such small changes, one probably need to perform calculations for more values of $y$ in order for the interpolation scheme used here to succeed. The changes in the average cation-anion bond distances with increasing Fe concentration can be explained from the results of Table \ref{table1}. Results of Table \ref{table1} suggest that the average bond distance at the tetrahedral sites decrease in presence of cation disorder due to the presence of Fe$^{3+}$ with the smallest ionic radii. Subsequent presence of Co$^{2+}$ at the octahedral sites increase the average bond distances. These lead to a decrease(increase) of the $A-O$($B-O$) bond distances with $x$. It can be noted that the various cation-anion bond distances at various sites are quite insensitive to the degree of cation disorder so that the qualitative variations of the $A-O$ and $B-O$ with $x$ can be explained purely as a composition effect.
The variations in the lattice parameters with $y$,  can be understood from the results of Table \ref{table1}.
The general trend seen in Table \ref{table2} is that with increase in $y$, the
lattice constants decrease, particularly for $x \geq 0.5$. This can be understood from the 
compositions of the tetrahedral and octahedral sites for different $y$ and $x$ and the variations
in their bond distances (Table \ref{table1}). For smaller $x$ values, the equilibrium value of $y$ is around 0.5. As a result, the percentages of Fe at A site and Co at B site are not too significant and the lattice parameter $a$, in particular is not affected by the changes in the degree of cation disorder. For $x \geq 0.5$, $y_{0} \geq 0.5$, and thus the $A-O$ bonds shorten considerably on an average effecting an overall decrease in the lattice parameters. 

\subsection{Magnetic exchange interactions and their dependencies on $x$ and $y$} 

\begin{figure}[ht]
\includegraphics[width=8cm,height=8cm]{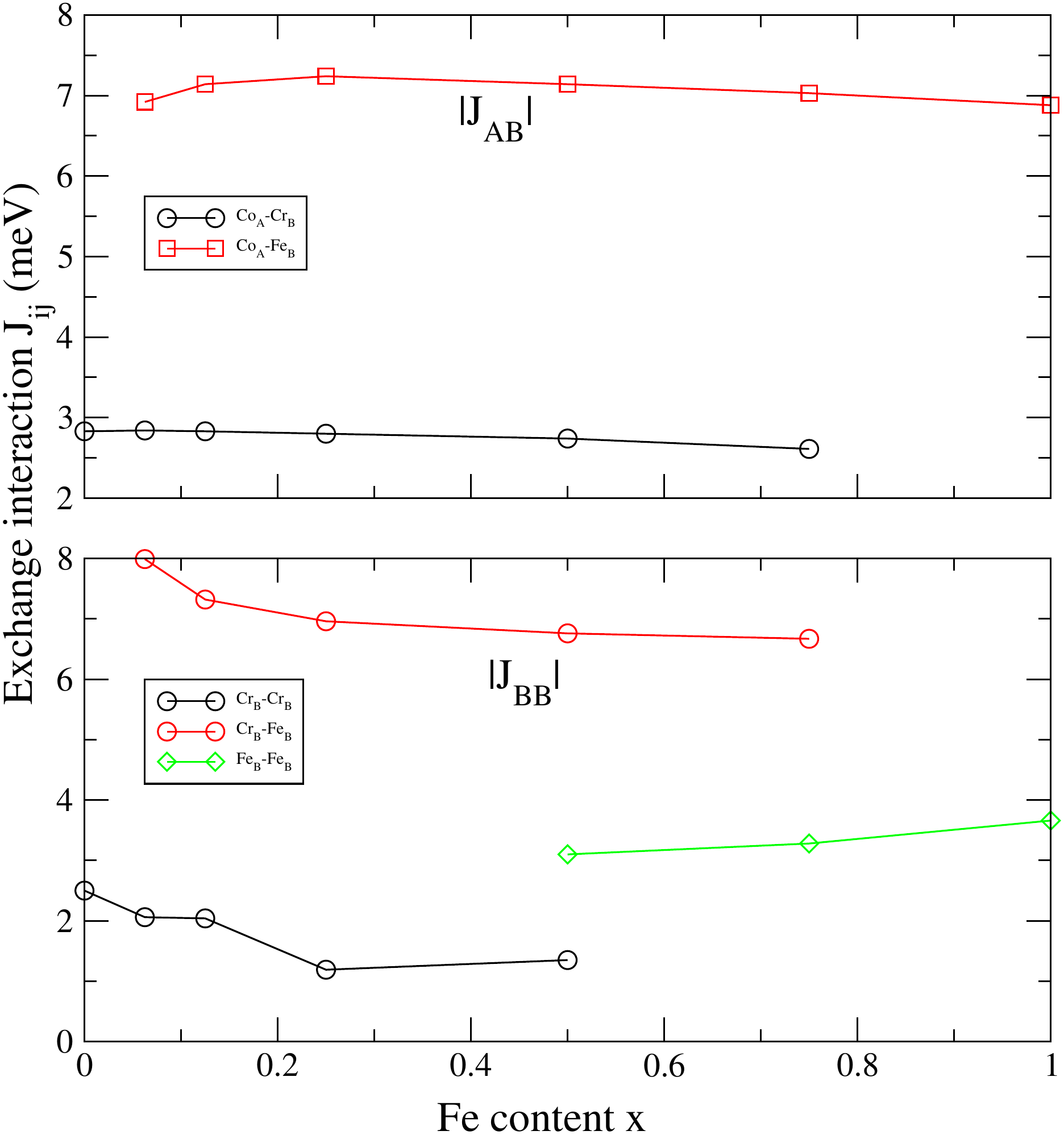}
\caption{\label{fig4} The absolute values of different magnetic exchange parameters calculated in the present work as a function of Fe concentration $x$ in Co$\left(Cr_{1-x}Fe_{x} \right)_{2}$O$_{4}$. The B-B exchange interactions(in meV) are indicated by $|J_{BB}|$ , while the A-B exchange interactions(in meV) are indicated by $|J_{AB}|$. The results are for the "normal" spinel structure, that is, for $y=0$.}
\end{figure}

\begin{figure}[ht]
\includegraphics[width=8cm,height=8cm]{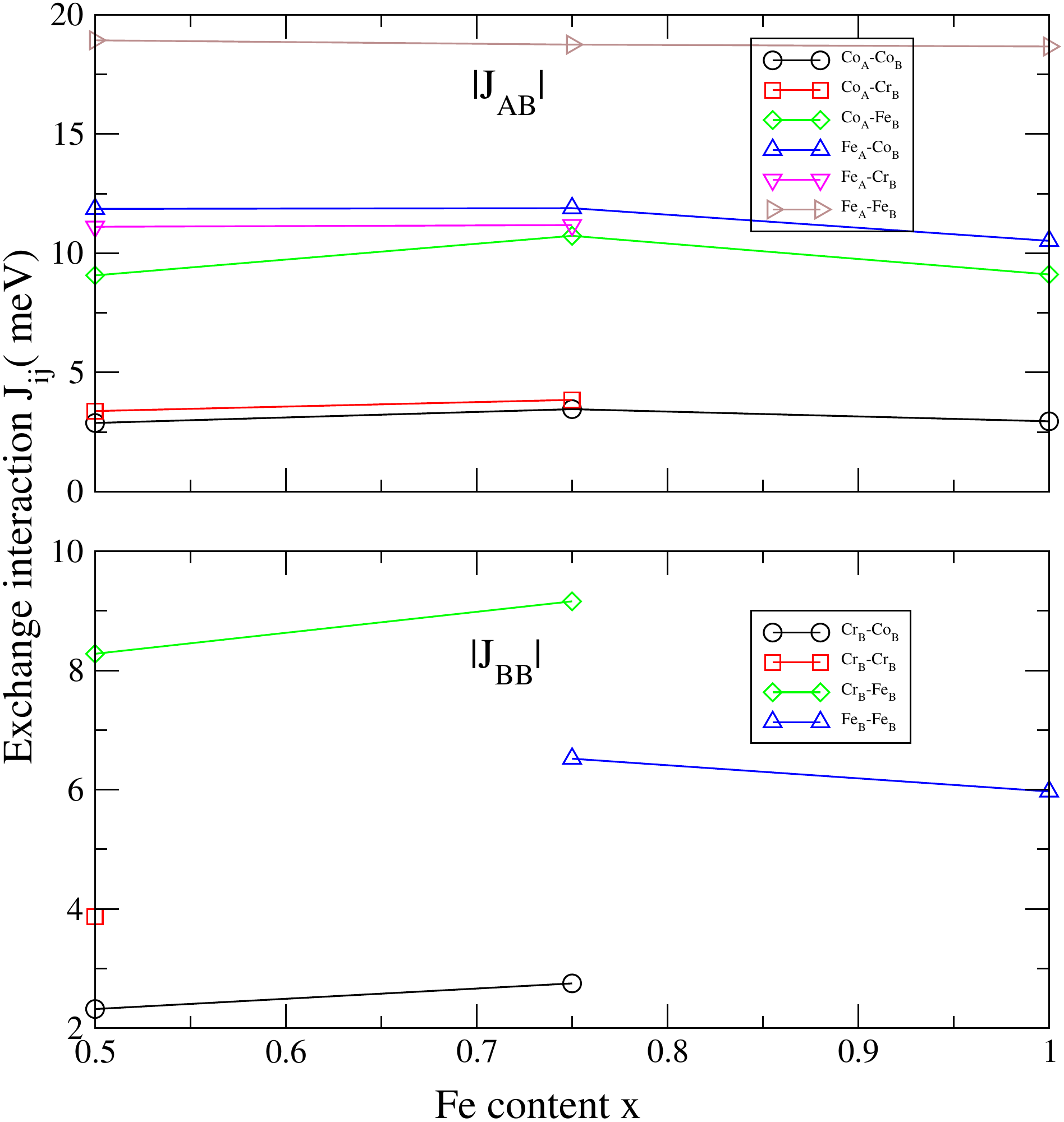}
\caption{\label{fig5} The absolute values of different magnetic exchange parameters calculated in the present work as a function of Fe concentration $x$ in Co$\left(Cr_{1-x}Fe_{x} \right)_{2}$O$_{4}$. The B-B exchange interactions(in meV) are indicated by $|J_{BB}|$ , while the A-B exchange interactions(in meV) are indicated by $|J_{AB}|$. The results are for the "half-inverse" spinel structure, that is, for $y=0.5$.}
\end{figure}

\begin{figure}[ht]
\includegraphics[width=8cm,height=8cm]{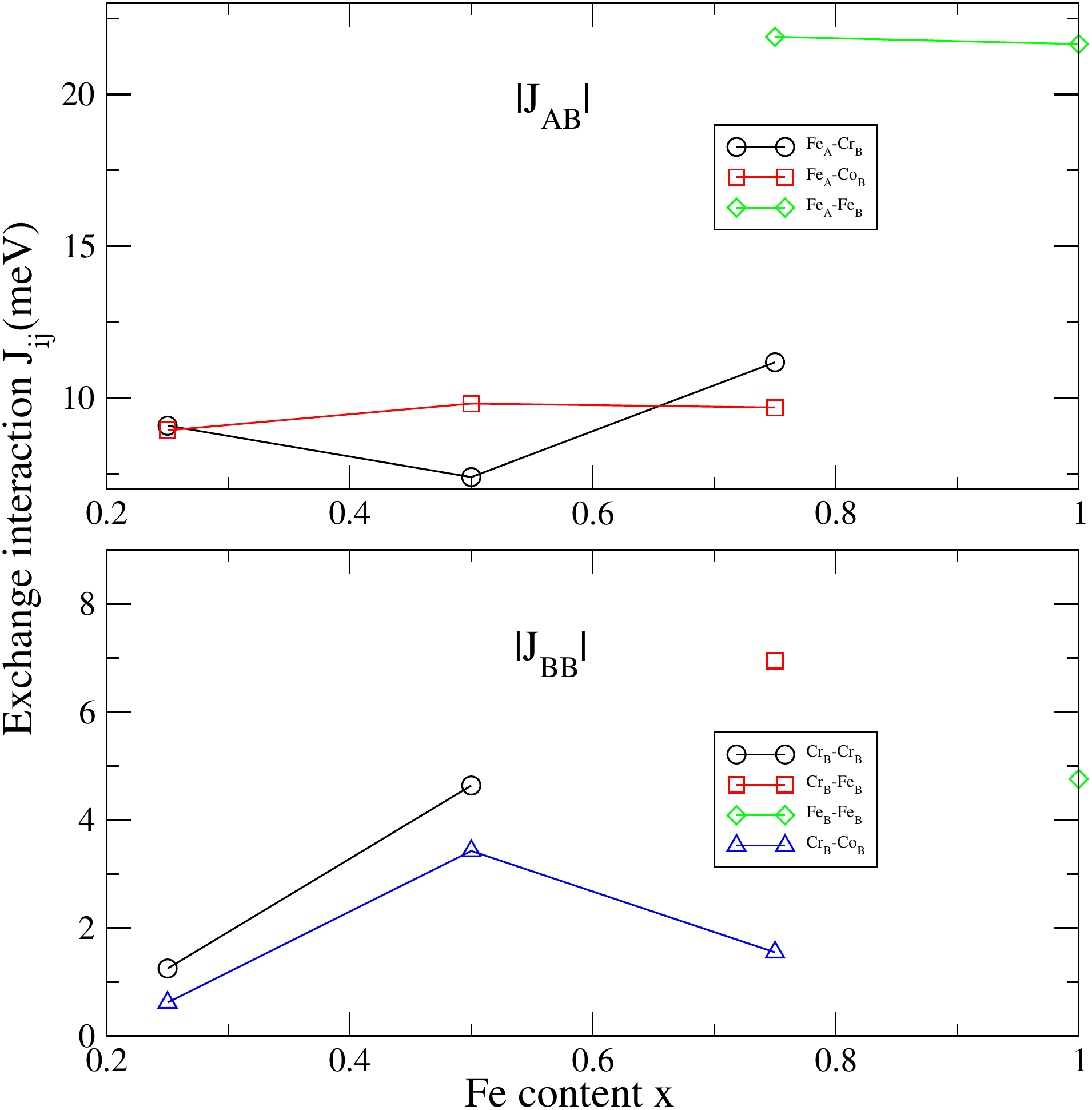}
\caption{\label{fig6} The absolute values of different magnetic exchange parameters calculated in the present work as a function of Fe concentration $x$ in Co$\left(Cr_{1-x}Fe_{x} \right)_{2}$O$_{4}$. The B-B exchange interactions(in meV) are indicated by $|J_{BB}|$ , while the A-B exchange interactions(in meV) are indicated by $|J_{AB}|$. The results are for the "complete inverse" spinel structure, that is, for $y=1$}
\end{figure}

 We now proceed to discuss how the Fe substitution in CoCr$_{2}$O$_{4}$ leads to a collinear magnetic structure. We are particularly interested in the evolution of the exchange interactions between different sites and different atoms as the composition of the system changes along with the degree of cation disorder.  As the nearest neighbour exchange interactions are sufficient to understand the trends \cite{bs}, we are interested in the nearest neighbour exchange interactions only. These are calculated by mapping the DFT+U total energies on a Heisenberg Hamiltonian \cite{bs,dd1,dd2}. Due to the requirement of prohibitively large resources for calculations of all exchange interactions, particularly for $y=0.5$ and $y=1$, we could compute certain exchange parameters only for a limited number of compositions. In spite of these limitations, the required trends to understand the evolution of the magnetic structure can be understood from the computed results as discussed in the following. In Reference \cite{bs}, evolution of the exchange interactions were computed by the Lichtenstein formulation \cite{lich} within the framework of KKR-CPA Green's function method \cite{sprkkr} which is free of the above mentioned limitations. However, the shortcoming of the KKR-CPA method is in their inability to treat the relaxations around atomic sites leading to local distortions and dispersions in bond distances. As a result, in Reference \cite{bs}, calculations could be done only for low values of $x(0 \leq x \leq 0.1)$. Their calculations with "normal" and "complete inverse" configurations showed that although there is a competition between the exchange interactions between the A and the B sites ($J_{AB}$) and the ones between B and B sites ($J_{BB}$), with increase in Fe content, the $J_{AB}$s increase rapidly while the $J_{BB}$s either decrease or increase slowly, indicating that $J_{AB}$s may superceed $J_{BB}$s with higher Fe content. In Figures \ref{fig4},\ref{fig5},\ref{fig6}, we present our calculated values of different $J_{ij}$ as a function of Fe content $x$ for three different degrees of cation disorder.  The results suggest that in case of Fe substitution at B sites, the "normal spinel" configuration ($y=0$), there is a competition between the exchange parameters associated with Co$_{A}$-Fe$_{B}$ (Co$_{A}$ stands for Co atoms at A sites, Fe$_{B}$ stands for Fe atoms at B sites) and Cr$_{B}$-Fe$_{B}$ pairs. Across the composition range, these two are the dominant interactions. At low values of $x$, the Cr$_{B}$-Fe$_{B}$ interaction is slightly dominant but the Co$_{A}$-Fe$_{B}$ interactions catch up as Fe concentration in the system increases. The Cr$_{B}$-Cr$_{B}$ and Co$_{A}$-Cr$_{B}$ interactions are relatively weak and slowly change with composition. Since the Fe atoms would like to distribute themselves homogeneously in the system \cite{dd3,bs}, Fe does not have another Fe as it's nearest neighbour until the concentration of Fe in the system is significant. This is demonstrated by the emergence of the Fe-Fe exchange parameter only at $x=0.5$. The strength of this interaction lies in between that of the dominant Cr$_{B}$-Fe$_{B}$ and weaker Cr$_{B}$-Cr$_{B}$ interactions, and would add to the strength of the total J$_{BB}$, thus competing closely with total J$_{AB}$ interactions. Thus, with the "normal" configuration, one would not achieve a collinear magnetic structure in this system. In Figs. \ref{fig5},\ref{fig6} we observe the effects of the cation disorder on the relative strengths of the exchange interactions. In the "half-inverse" configuration ($y=0.5$), the A-B interactions overwhelmingly dominate over the B-B interactions for $0.5 \leq x \leq 1$. It may be noted that modelling of a "half-inverse" structure for $x < 0.5$ would have required larger unit cells and hence we refrained from modelling the "inverse" structures for smaller $x$ values. This would not pose any serious problem as we are interested only to find the trends in different exchange interactions when the state of cation disorder changes, in order to understand the origin of the emergence of the collinear structure at one end. Analysing Fig. \ref{fig5}, we further see that the dominant A-B interaction for $y=0.5$ comes from the Fe atoms at crystallographic sites of different symmetry. The three other dominant A-B exchange interactions Co$_{A}$-Fe$_{B}$,Fe$_{A}$-Co$_{B}$ and Fe$_{A}$-Cr$_{B}$ compete with each other. On the other hand, the dominant B-B interaction is between Fe and Cr atoms at B sites as was the case for $y=0$. The picture remains same for $y=1$, that is, for "complete inverse" configuration. Thus, it can be concluded that in Co$\left(Cr_{1-x}Fe_{x} \right)_{2}$O$_{4}$, a collinear magnetic structure would emerge due to the cation disorder among A and B sites, and would certainly be achieved when Fe composition 
 would reach $50 \%$.   
The question now remains as to why a particular magnetic exchange parameter dominates over the others and how that changes with changes in the degree of cationic disorder. In the next section, we address this question by analysing the electronic structures as a function of composition and degree of cation disorder.

 \subsection{ Electronic structures and their variations with $x$ and $y$}
\begin{figure}[!t]
\includegraphics[width=8.0cm,height=10.0cm]{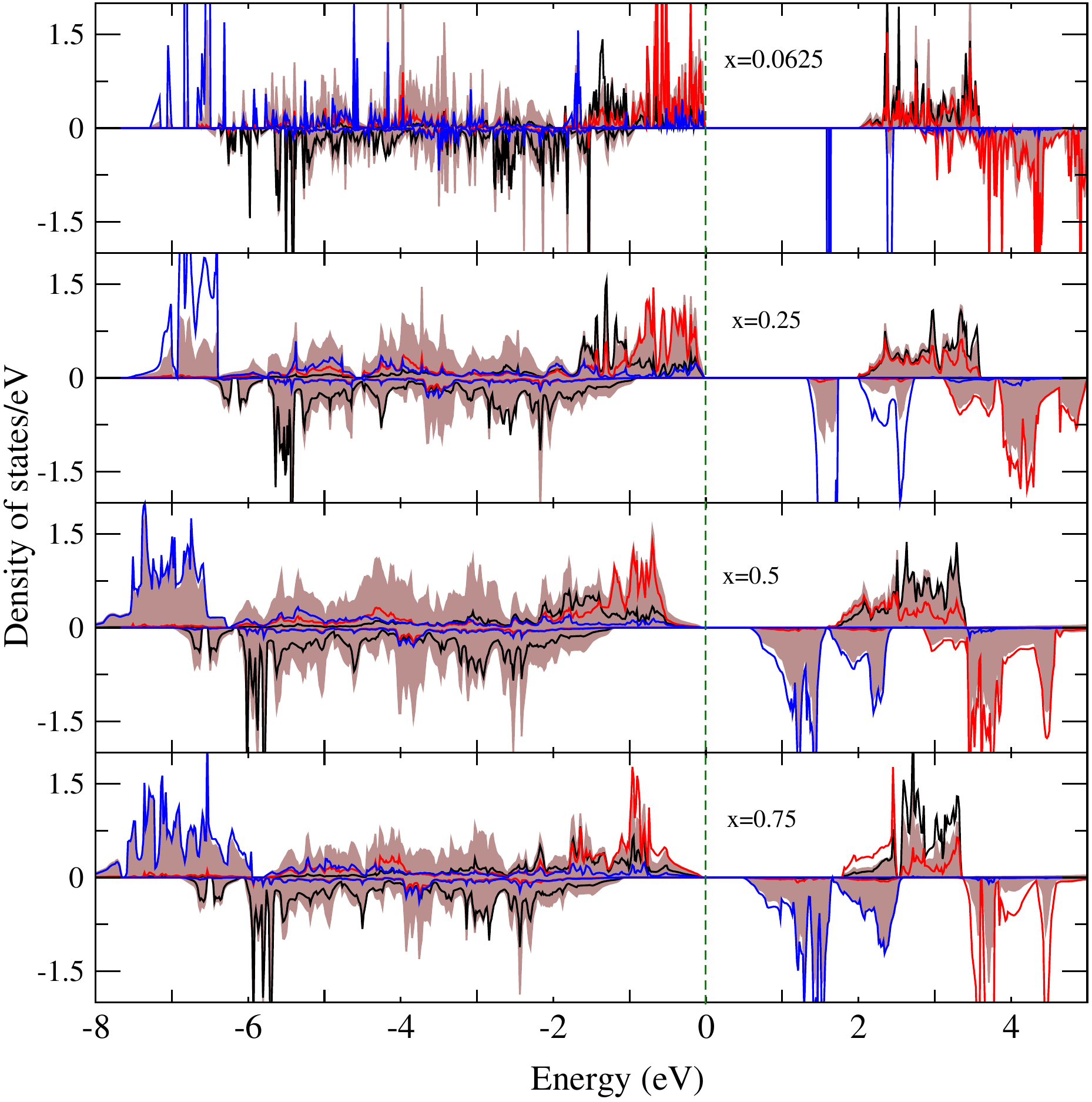}
\caption{\label{fig7} Total and atom-projected densities of states of 
$Co\left(Cr_{1-x}Fe_{x} \right)_{2}O_{4}$
for different $x$. The results are for zero cation disorder ($y=0$). Here  the total densities of states is denoted by brown shade. The black, red and blue
curves represent atom projected densities of states of Co at tetrahderal
sites, Cr and Fe at octahedral sites, respectively.}
\end{figure}
\begin{figure}[!t]
\includegraphics[width=8.0cm,height=10.0cm]{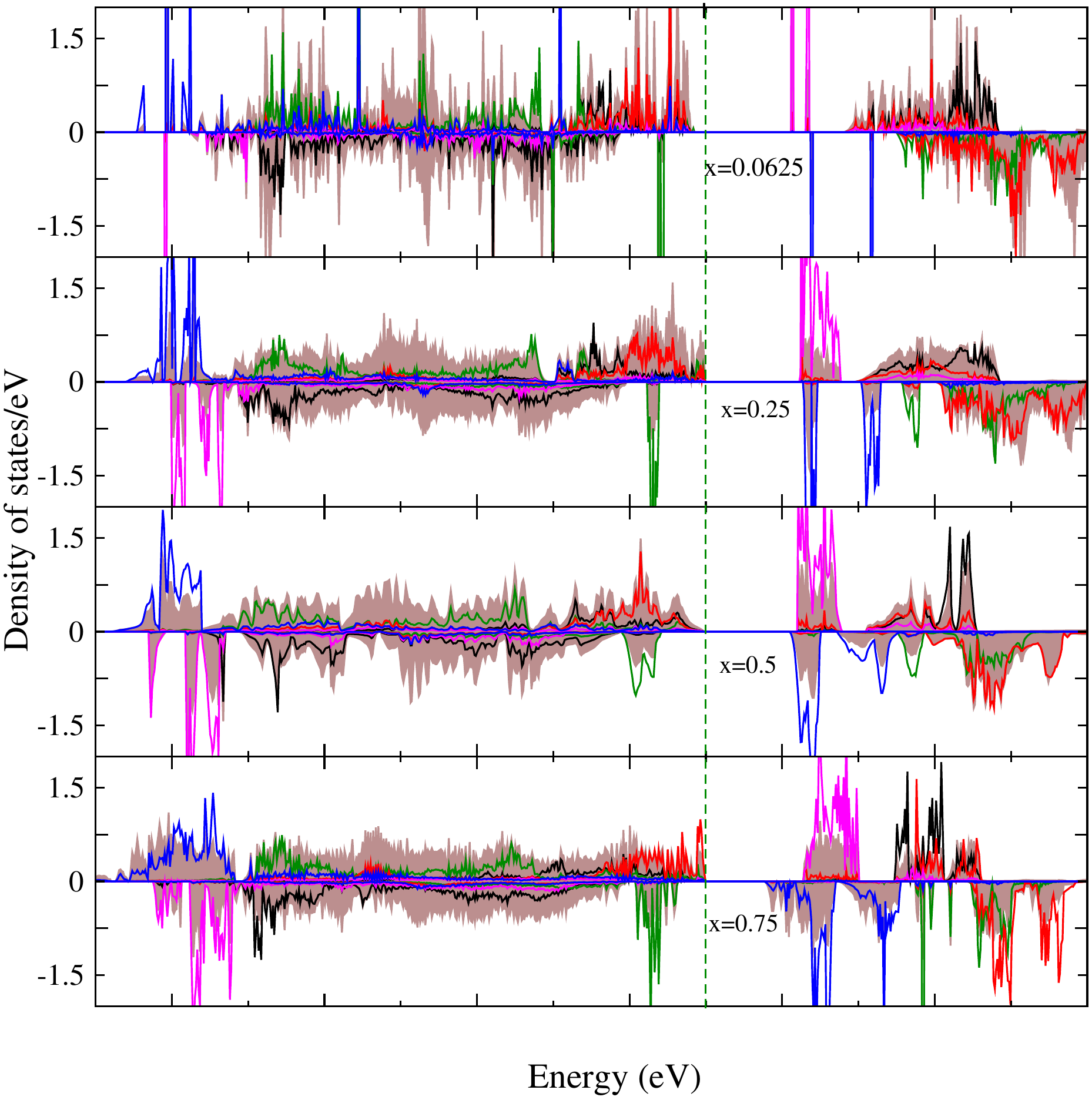}
\caption{\label{fig8} Total and atom-projected densities of states of 
$Co\left(Cr_{1-x}Fe_{x} \right)_{2}O_{4}$
for different $x$. The results are for $50 \%$ cation disorder ($y=0.5$).Here          
 the total densities of states is denoted by brown shade. The black and the 
green curves 
represent atom projected densities of states for Co at tetrahedral and 
at octahedral sites respectively, the red curve represents atom
projected densities of states for Cr, the pink and the blue curves represent atom
projected densities of states for Fe atoms at tetrahedral and at octahedral sites
respectively.}
\end{figure}
\begin{figure}[!t]
\includegraphics[width=8.0cm,height=10.0cm]{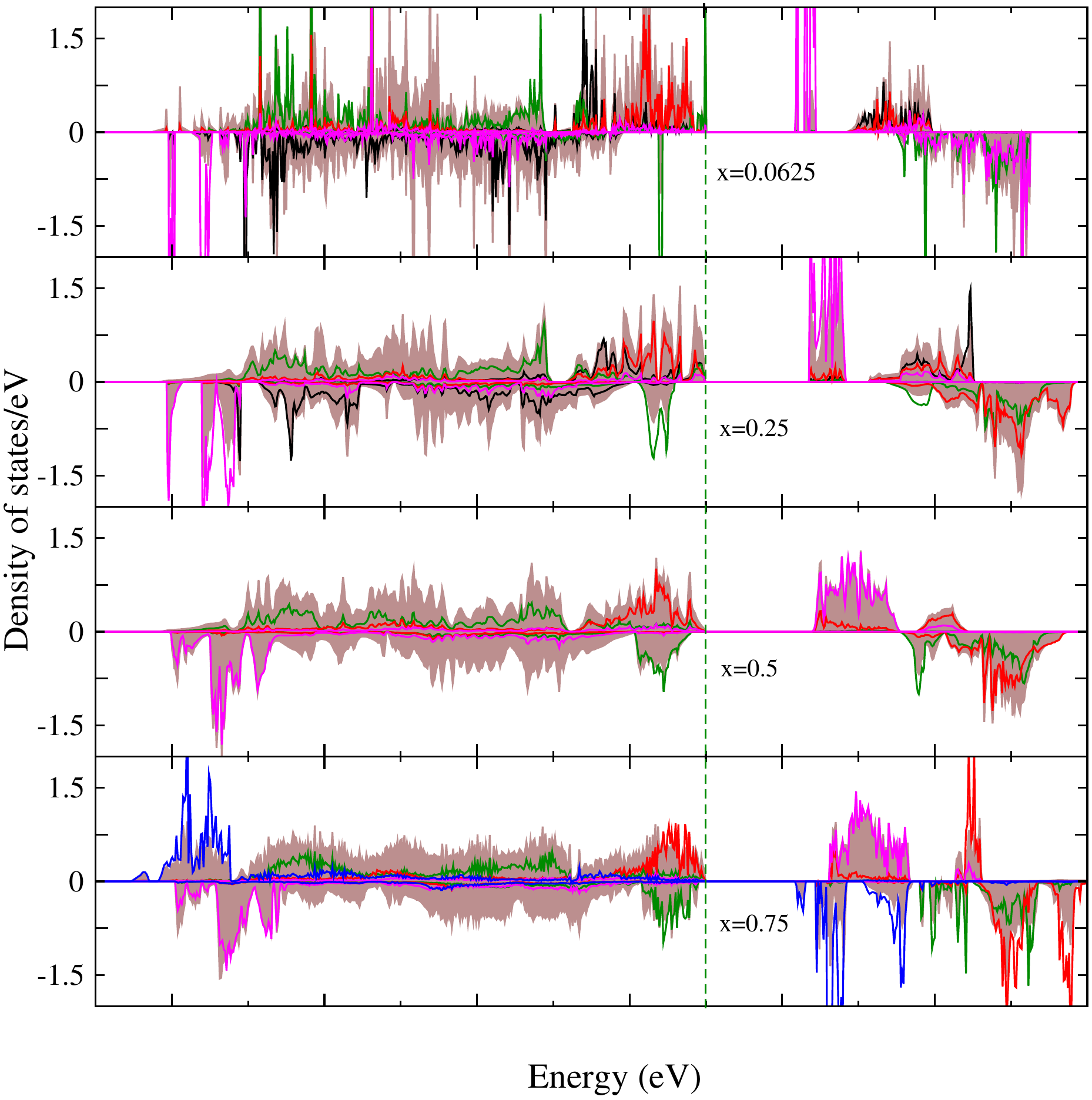}
\caption{\label{fig9}Total and atom-projected densities of states of                
$Co\left(Cr_{1-x}Fe_{x} \right)_{2}O_{4}$
for different $x$. The results are for full inverse arrangement ($y=1$).Here
 the total densities of states is denoted by brown shade. The black and the 
green curves
represent atom projected densities of states for Co at tetrahedral and 
at octahedral sites respectively, the red curve represents atom
projected densities of states for Cr, the pink and the blue curves represent atom
projected densities of states for Fe atoms at tetrahedral and at octahedral sites
respectively.}  
 \end{figure}
We present results on the electronic structures with variations in $x$ for three
degrees of cation disorder represented by $y=0,0.5$ and $1$ in Figures \ref{fig7},\ref{fig8},\ref{fig9} 
respectively. Quite interestingly, the qualitative features in the atom-projected densities of states remain absolutely same irrespective of $x$ and $y$. A careful look at the figures suggest the following: for any $x$ and $y$, Co$_{A}$ t$_{2g}$ states are  half-filled and the  $e_{g}$ states are completely filled, Cr t$_{2g}$ states are half-filled while e$_{g}$ states are completely empty, Fe states at both A and B sites have both t$_{2g}$ and e$_{g}$ states half-filled while Co$_{B}$ t$_{2g}$ states are more than half-filled and the e$_{g}$ states are half-filled. Thus, the band filling associated with the atoms resemble that of pure CoCr$_{2}$O$_{4}$ and CoFe$_{2}$O$_{4}$ \cite{dd2}. Such robustness of the electronic structures originates from strong and comparable crystal fields associated with the magnetic cations as was seen in cases of pure CoCr$_{2}$O$_{4}$ and CoFe$_{2}$O$_{4}$ \cite{dd2}. The positions of the major peaks associated with each individual atoms also do not change their positions considerably with changes in $x$ and $y$. The only change that one can see is the re-distributions of the weights associated with a particular atom type when $x$ changes, keeping $y$ fixed. 

The features in the densities of states can now be conveniently used to explain the trends in the magnetic exchange interactions. For $y=0$, the results presented in Fig. \ref{normalxc} show that Co$_{A}$-Fe$_{B}$ exchange interaction closely compete with Cr$_{B}$-Fe$_{B}$ one. Since Co$_{A}$  and Fe$_{B}$ have half-filled t$_{2g}$ and e$_{g}$ states respectively, this interaction is strongly antiferromagnetic \cite{wickham}. On the other hand, both Cr$_{B}$ and Fe$_{B}$ t$_{2g}$ orbitals are half-filled. Direct interaction between them is possible which strengthens the overall B-B interaction \cite{wickham}. As a result, there is close competition between $J_{AB}$ and $J_{BB}$ exchange interactions for $y=0$. The strength of the dominant $J_{BB}$, however, is weaker in comparison to the dominant $J_{AB}$ due to the fact that the inter-cation distance at the B site is larger than that at the A site as can be understood from the bond distances presented in Table \ref{table1}. For $y=0.5,1$, the overwhelmingly dominant A-B interaction is due to the Fe atoms at the A and the B sites. In this case, both A site t$_{2g}$ and B site e$_{g}$  orbitals are half-filled making the interaction strongly antiferromagnetic.  The relatively much weaker B-B interaction, mostly coming from Cr-Fe pairs are due to the fact that due to inversion, the Fe content at B site decreases in comparison to the $y=0$ case, thus, increasing the cation-cation distances at the B site. The other prominent A-B interactions for $y=0.5,1$ are due to Fe$_{A}$-Co$_{B}$ and Fe$_{A}$-Cr$_{B}$ pairs. In case of the former, although the A site t$_{2g}$ and B site e$_{g}$ are half-filled, and thus the exchange interaction should be strongly antiferromagnetic, they are weaker in comparison to the Fe-Fe interactions. This is due to the fact that Fe$_{A}$-Co$_{B}$ distances are larger than the Fe$_{A}$-Fe$_{B}$ distances (Table \ref{table1}). In case of Fe$_{A}$-Cr$_{B}$ pair, the empty e$_{g}$ orbitals at the Cr sites make the interactions relatively weak \cite{wickham}.   
 
\section{Summary and Conclusions}
We have explored the effects of cation disorder on the structural and magnetic properties of Co$\left(Cr_{1-x}Fe_{x} \right)_{2}$O$_{4}$ by computing the "inversion parameter" at each value of $x$ from a combination of DFT based total energy calculations and a thermodynamic model. In agreement with the inferences from the experimental results on structural, magnetic and thermal properties of this system, we find that for all compositions, Fe has a strong preference for the tetrahedral A sites in the spinel structure. We have explicitly calculated the structural parameters as a function of the degree of inversion and the composition. The qualitative trends in our results agree very well with the experimentally observed ones. The major highlight of this work is that we have provided an explanation to the emergence of a collinear spin structure from a non-collinear one as the Fe substitution in the system increases gradually. We have conclusively shown that the degree of cation disorder is the deciding factor for the emergence of a collinear magnetic order in this system. The trends in the magnetic exchange interactions have been explained from the features in the electronic structures. In comparison to Mn substituted CoCr$_{2}$O$_{4}$, the features in the electronic structures of Fe substituted CoCr$_{2}$O$_{4}$ are far more robust against changes in the composition and the degree of cation disorder. This can be understood from the relatively strong and comparable crystal fields associated with the cations at different sites of the end compounds. This work, therefore, stands out as the first comprehensive investigations into the fundamentals of the Fe substituted CoCr$_{2}$O$_{4}$ for a larger composition regime than what has been available so far.

\section{Acknowledgments}
The computation facilities from C-DAC, Pune, India
and from Department of Physics, IIT Guwahati funded under the FIST programme of DST, India
are acknowledged.

\end{document}